\documentclass[aps,prl,preprint]{revtex4}

\begin{document}

\title{On the two-photon contributions to $e^+e^-\to\eta\gamma$
and $e^+e^-\to\eta^\prime\gamma$}
\author{ Z.~K.~Silagadze} 
\affiliation{Budker Institute of Nuclear Physics and Novosibirsk State 
University, 630 090, Novosibirsk, Russia }

\begin{abstract}
Motivated by recent BABAR measurements of the $\gamma^*\to\gamma \eta$ and
$\gamma^*\to\gamma \eta^\prime $ transition form factors, we estimate 
two-photon exchange contributions to the corresponding cross sections.
By using a phenomenological model, based on the vector meson dominance, it
is argued that the expected contributions are small enough not to effect the
BABAR results. As a by product we predict $Br(\eta^\prime\to\mu^+\mu^-)=
(1.4\pm 0.2)\times 10^{-7}$. Our results might be useful also in high 
precision calculations of radiative corrections to the Dalitz decays of 
pseudoscalar mesons.
\end{abstract}

\maketitle

\section{Introduction}
Recently, the BABAR collaboration reported \cite{1-1} measurements of the
$e^+e^-\to\eta\gamma$ and $e^+e^-\to\eta^\prime\gamma$ cross sections
and the corresponding $\gamma^*\to\gamma P$ transition form factors
at a center-of-mass energy $\sqrt{s}=10.58~\mathrm{GeV}$. In the asymptotic
limit of large photon virtualities these form factors are determined by
perturbative QCD in terms of the decay constants of pseudoscalar mesons.
However, the comparison of the BABAR result with the QCD predictions
is hampered by the fact that the determination of the $\eta$ and $\eta^\prime$ 
decay constants requires taking into account the mixing between these two
states and different phenomenological models, cited in \cite{1-1}, give
different results. Amusingly, different models are needed to reconcile
asymptotic QCD predictions for the $\eta$ and $\eta^\prime$ transition 
form factors with the BABAR result and neither model can explain the 
measured ratio of these form factors:  $1.10\pm 0.17$ versus theoretical
predictions in the range from $1.6$ to $2.3$. This discrepancy, first of all,
calls for a careful examination of possible sources of unaccounted background. 

\input FEYNMAN
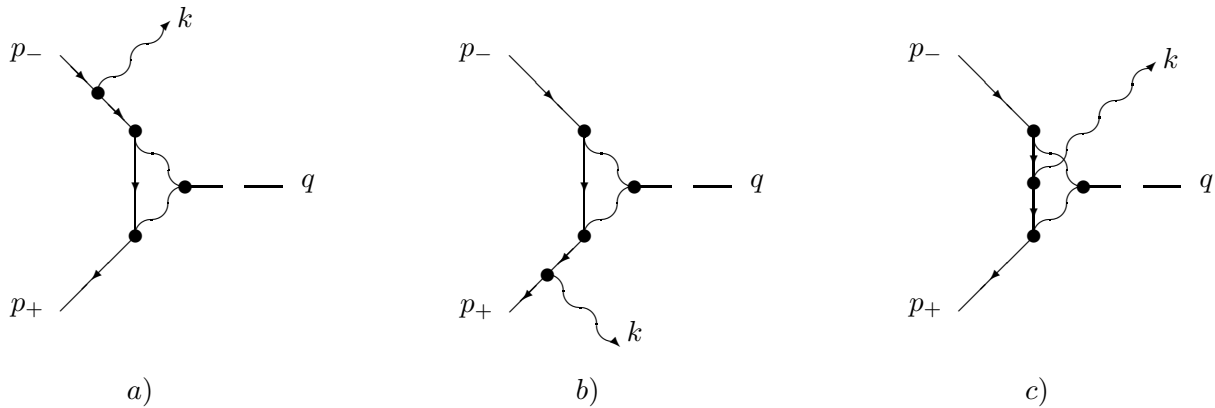
\begin{figure}[htp]
\begin{center}
\begin{picture}(46000,15000)
\drawline\fermion[\SE\REG](2000,13000)[2000]
\drawarrow[\SE\ATBASE](\pmidx,\pmidy)
\global\advance\pfrontx by -1900
\put(\pfrontx,\pfronty){$p_-$}
\put(\pbackx,\pbacky){\circle*{500}}
\drawline\photon[\NE\REG](\pbackx,\pbacky)[4]
\drawarrow[\NE\ATBASE](\pbackx,\pbacky)
\global\advance\pbackx by 500
\put(\pbackx,\pbacky){$k$}
\drawline\fermion[\SE\REG](\pfrontx,\pfronty)[2000]
\drawarrow[\SE\ATBASE](\pmidx,\pmidy)
\put(\pbackx,\pbacky){\circle*{500}}
\drawline\fermion[\S\REG](\pbackx,\pbacky)[4000]
\drawarrow[\S\ATBASE](\pmidx,\pmidy)
\put(\pbackx,\pbacky){\circle*{500}}
\drawline\fermion[\SW\REG](\pbackx,\pbacky)[4000]
\global\advance\pbackx by -1900
\put(\pbackx,\pbacky){$p_+$}
\drawarrow[\SW\ATBASE](\pmidx,\pmidy)
\drawline\photon[\NE\REG](\pfrontx,\pfronty)[3]
\drawline\scalar[\E\REG](\photonbackx,\photonbacky)[2]
\global\advance\pbackx by 700
\put(\pbackx,\pbacky){$q$}
\put(\pfrontx,\pfronty){\circle*{500}}
\drawline\photon[\NW\REG](\photonbackx,\photonbacky)[3]
\put(4500,0){$a)$}
\drawline\fermion[\SE\REG](19000,13000)[4000]
\drawarrow[\SE\ATBASE](\pmidx,\pmidy)
\global\advance\pfrontx by -1900
\put(\pfrontx,\pfronty){$p_-$}
\put(\pbackx,\pbacky){\circle*{500}}
\drawline\fermion[\S\REG](\pbackx,\pbacky)[4000]
\drawarrow[\S\ATBASE](\pmidx,\pmidy)
\put(\pbackx,\pbacky){\circle*{500}}
\drawline\photon[\NE\REG](\pbackx,\pbacky)[3]
\drawline\scalar[\E\REG](\photonbackx,\photonbacky)[2]
\global\advance\pbackx by 700
\put(\pbackx,\pbacky){$q$}
\put(\pfrontx,\pfronty){\circle*{500}}
\drawline\photon[\NW\REG](\photonbackx,\photonbacky)[3]
\global\advance\pbacky by -3800
\drawline\fermion[\SW\REG](\pbackx,\pbacky)[2000]
\drawarrow[\SW\ATBASE](\pmidx,\pmidy)
\put(\pbackx,\pbacky){\circle*{500}}
\drawline\photon[\SE\REG](\pbackx,\pbacky)[4]
\drawarrow[\SE\ATBASE](\pbackx,\pbacky)
\global\advance\pbackx by 500
\put(\pbackx,\pbacky){$k$}
\drawline\fermion[\SW\REG](\pfrontx,\pfronty)[2000]
\drawarrow[\SW\ATBASE](\pmidx,\pmidy)
\global\advance\pbackx by -1900
\put(\pbackx,\pbacky){$p_+$}
\put(21500,0){$b)$}
\drawline\fermion[\SE\REG](36000,13000)[4000]
\drawarrow[\SE\ATBASE](\pmidx,\pmidy)
\global\advance\pfrontx by -1900
\put(\pfrontx,\pfronty){$p_-$}
\put(\pbackx,\pbacky){\circle*{500}}
\drawline\fermion[\S\REG](\pbackx,\pbacky)[2000]
\put(\pbackx,\pbacky){\circle*{500}}
\drawarrow[\S\ATBASE](\pmidx,\pmidy)
\drawline\photon[\NE\REG](\pbackx,\pbacky)[7]
\drawarrow[\NE\ATBASE](\pbackx,\pbacky)
\global\advance\pbackx by 500
\put(\pbackx,\pbacky){$k$}
\drawline\fermion[\S\REG](\pfrontx,\pfronty)[2000]
\drawarrow[\S\ATBASE](\pmidx,\pmidy)
\put(\pbackx,\pbacky){\circle*{500}}
\drawline\fermion[\SW\REG](\pbackx,\pbacky)[4000]
\global\advance\pbackx by -1900
\put(\pbackx,\pbacky){$p_+$}
\drawarrow[\SW\ATBASE](\pmidx,\pmidy)
\drawline\photon[\NE\REG](\pfrontx,\pfronty)[3]
\drawline\scalar[\E\REG](\photonbackx,\photonbacky)[2]
\global\advance\pbackx by 700
\put(\pbackx,\pbacky){$q$}
\put(\pfrontx,\pfronty){\circle*{500}}
\drawline\photon[\NW\REG](\photonbackx,\photonbacky)[3]
\put(38500,0){$c)$}
\end{picture}
\end{center}
\caption{Two-photon contribution to the $e^+e^-\to P\gamma$ reaction.}
\label{fig1-1}\end{figure}

Two-photon contribution to the $e^+e^-\to P\gamma$ reaction, described by
the diagrams shown in Fig.\ref{fig1-1}, is one such potential background 
source. As was mentioned in \cite{1-1}, this background is expected to be 
very small. But no detailed calculations exist in the literature, to our
knowledge, to support this conclusion. In this paper, we will try to fill up 
this gap. 

Naively one can think that this contribution shares the $(m_e/M_P)$ 
suppression factor, inherent to the $P\to e^+ e^-$ decay, because the graphs
in Fig.\ref{fig1-1} are obtained from $e^+ e^-\to\gamma^*\gamma^*\to P$
by the insertion of a bremsstrahlung photon on the electron line. But this 
is not the case as was realized long ago \cite{1-2,1-3,1-4} in the context 
of the $P\to e^+ e^- \gamma$ decay.

As we see from Fig.\ref{fig1-1}, off-shell $P\to\gamma^*\gamma^*$ amplitude 
is needed to calculate the two-photon contribution to the $e^+e^-\to P\gamma$ 
reaction. Therefore we first discuss this amplitude.

\section{Vector Meson Dominance model}
We define $\eta(Q)\to\gamma^*(k_1,\mu)\,\gamma^*(k_2,\nu)$ amplitude 
$A_{\mu\nu}$ through the $F_{\eta\gamma^*\gamma^*}$ form factor 
as follows
\begin{equation}
A_{\mu\nu}=-i\frac{e^2}{f_\pi}F_{\eta\gamma^*\gamma^*}(k_1^2,k_2^2)
\,\epsilon_{\mu\nu\sigma\tau}k_1^\sigma k_2^\tau .
\label{eq2-1} \end{equation}
The pion decay constant $f_\pi\approx 93~\mathrm{MeV}$ is introduced
to make the form factor dimensionless. For on-shell photons and in the 
chiral limit, $Q^2=0$, the $F_{\eta\gamma^*\gamma^*}$ form factor is fixed 
by chiral anomaly \cite{2-1,2-2,2-3,2-4,2-5}
$$F_{\eta\gamma^*\gamma^*}(0,0)=\frac{\alpha}{\sqrt{3}\,\pi},$$ 
while for large photon virtualities QCD predicts
\cite{2-6,2-7,2-8,2-9} $\sim 1/k^2$ asymptotic behaviour.

However, in the interim region the fundamental theories tell little about
the form factor behaviour and one has to rely on phenomenological models
like the Vector Meson Dominance (VMD) model \cite{2-10,2-11}. Historically 
this model emerged in attempts to understand photon interactions with 
hadronic matter and in its extreme form asserts that hadronic interactions
of the photon proceed exclusively through known vector mesons. At low energy
region of the lightest vector mesons ($\rho$, $\omega$ and $\phi$) VMD was
extremely successful in describing a wide range of experimental data 
\cite{2-12}. This success naturally rises a question whether there is 
a theoretical justification of the VMD from modern perspective of the 
Standard Model.

One of the cornerstones of the Standard Model is gauge principle. The naive
$\gamma-\rho$ direct coupling is not compatible with 
gauge invariance as it leads to the photon acquiring an imaginary mass.
Nevertheless it was shown by Kroll, Lee, and Zumino \cite{2-13} that the
complete vector meson dominance is consistent with gauge invariance provided 
these mesons are coupled only to conserved currents. In phenomenological
VMD based applications an additional terms in the interaction Lagrangian
implied by the Kroll-Lee-Zumino analysis (for example a photon mass term)
are usually neglected as they are of higher order in $\alpha$.

Another guiding principle of the Standard Model is symmetry. At low energies
we cannot solve QCD effectively  but its symmetries still guide us in 
constructing effective theories of colorless hadrons, which are the only 
relevant QCD degrees of freedom in this non-perturbative region. Pseudoscalar
mesons play a special role in this game as they are Goldstone bosons 
associated with the spontaneous breaking of QCD chiral symmetry (in fact 
would-be Goldstone bosons because the chiral symmetry is explicitly broken by 
nonzero quark masses). Therefore their low energy interactions, encoded into
a chiral effective Lagrangian, are uniquely determined from symmetry 
considerations in terms of a few phenomenological parameters like the pion
decay constant $f_\pi$. The resulting Chiral Perturbation Theory (CHPT) 
is commonly considered nowadays as the effective field theory of the Standard
Model at low energies (for review see, for example, \cite{2-14,2-15,2-16}).

Although the general method how to add vector mesons (and any other 
particles) to chiral Lagrangians was formulated long ago \cite{2-17,2-18},
usually some dynamical principle is needed to reduce number of the free
parameters and enhance predictability of the theory. Hidden Local Symmetry 
(HLS) approach \cite{2-19,2-20} is considered to be the most convenient 
scheme to deal with vector mesons. Hidden local symmetries were initially 
discovered in the supergravity theories and after realizing that they are a 
common feature of any nonlinear sigma model Bando et al. suggested to 
consider vector mesons as dynamical gauge bosons of hidden local symmetry of 
nonlinear chiral Lagrangian \cite{2-19,2-20}. There is nothing special
about this ``hidden symmetry''. Simply it is a language that makes power 
counting in derivative expansion more convenient when vector mesons are 
light \cite{2-21}.
 
When electromagnetism is introduced in the HLS Lagrangian \cite{2-22} one 
finds that generally there is a direct coupling of photons to charged 
pseudoscalars. Only for particular choice of parameters,
\begin{equation}
\left (\frac{f_\pi g_{\rho\pi\pi}}{M_\rho}\right )^2=\frac{1}{2},
\label{eq2-2}\end{equation}
one recovers the complete vector meson dominance. Equation (\ref{eq2-2}) is
the celebrated Kawarabayashi-Suzuki-Riazuddin-Fayyazuddin (KSRF) relation
\cite{2-23}. Therefore, from this perspective, VMD is not fundamental 
derivative of the Standard Model but rather just a lucky dynamical accident 
of the three flavour QCD \cite{2-24}.

Nevertheless the KSRF relation (\ref{eq2-2}) and hence VMD is experimentally 
well satisfied. Therefore in our estimates we will use VMD form factors.
We do not assume $SU(3)$ relations between various coupling constants except
relative phases, but determine their magnitudes from phenomenology. The 
relevant coupling constants are defined as follows.

Each photon-vector meson vertex gives a $-ieg_{V\gamma}M_V^2$ factor in the
matrix element. The coupling constants $g_{V\gamma}$ are assumed to be 
positive and can be determined from the electronic widths 
$\Gamma(V\to e^+e^-)$:
$$\Gamma(V\to e^+e^-)=\frac{4\pi\alpha^2}{3}M_Vg^2_{V\gamma}\left (1+2\frac
{m_e^2}{M_V^2}\right )\sqrt{1-\frac{4m_e^2}{M_V^2}}\approx 
\frac{4\pi\alpha^2}{3}g^2_{V\gamma}M_V.$$ Using the PDG data \cite{2-25},
we get
\begin{equation}
g_{\rho\gamma}=0.2014\pm 0.0016,\;\;g_{\omega\gamma}=0.0586\pm 0.0010,\;\;
g_{\phi\gamma}=0.0747\pm 0.0012.
\label{eq2-3}\end{equation}
The $\eta\to V_\mu(p) V_\nu(q)$ transition gives a factor
$$-i\varepsilon_V \frac{g_{\eta VV}}{f_\pi}\epsilon^{\mu\nu\sigma\tau}
p_\sigma q_\tau.$$
This defines dimensionless positive constants $g_{\eta VV}$. Here 
$\varepsilon_{\rho,\omega}=1$ and $\varepsilon_\phi=-1$, which together with
the definition of the $g_{V\gamma}$ constants corresponds to relative phases
expected from the $SU(3)$ symmetry with standard mixing angles. Neglecting 
$g_{\eta\phi\omega}$, which vanishes for the ideal $\phi-\omega$ mixing 
and nonet symmetry and hence is expected to be small, other coupling 
constants can  be determined from the $\Gamma(V\to\eta\gamma)$ decay widths 
assuming VMD:
$$\Gamma(V\to\eta\gamma)=\frac{\alpha}{24}\,g^2_{\eta VV}g^2_{V\gamma}
\left (\frac{M_V}{f_\pi}\right )^2\left (1-\frac{M_\eta^2}{M_V^2}\right )^3
M_V.$$
The results are
\begin{equation}
g_{\eta\rho\rho}=0.723\pm 0.067,\;\; g_{\eta\omega\omega}=0.735\pm 0.054,
\;\; g_{\eta\phi\phi}=0.858\pm 0.019.
\label{eq2-4}\end{equation}
Now VMD completely determines the $F_{\eta\gamma^*\gamma^*}$ form factor
in terms of the above given coupling constants:
\begin{equation}
F_{\eta\gamma^*\gamma^*}(k_1^2,k_2^2)=\frac{g_{\eta\rho\rho}\,
g^2_{\rho\gamma}M^4_\rho}{(k_1^2-M_\rho^2)(k_2^2-M_\rho^2)}+
\frac{g_{\eta\omega\omega}\,g^2_{\omega\gamma}M^4_\omega}{(k_1^2-M_\omega^2)
(k_2^2-M_\omega^2)}-\frac{g_{\eta\phi\phi}\,g^2_{\phi\gamma}M^4_\phi}
{(k_1^2-M_\phi^2)(k_2^2-M_\phi^2)}.
\label{eq2-5}\end{equation}
As an immediate check, one can calculate  two photon decay width 
$\Gamma(\eta\to 2\gamma)$ using this form factor. It is convenient to express 
the result as the following sum rule
\begin{equation} 
\Gamma(\eta\to 2\gamma)=\frac{9M_\eta^3}{2\alpha}\left [ \sum
\limits_{V=\rho,\omega,\phi}\varepsilon_V\,\frac{M_V}
{M_V^2-M_\eta^2}\,\sqrt{\frac{\Gamma(V\to\eta\gamma)\,
\Gamma(V\to e^+e^-)}{M_V^2-M_\eta^2}}\;\right ]^2.
\label{eq2-6}\end{equation}
This relation, which  remains valid even for non-zero $g_{\eta\phi\omega}$,
is well satisfied experimentally: it gives $\Gamma(\eta\to 2\gamma)=
(0.582\pm 0.085~\mathrm{keV})$, while the experimental width is \cite{2-25}
$\Gamma(\eta\to 2\gamma)=(0.510\pm 0.026)~\mathrm{keV}$. 

For $\eta^\prime$ couplings, we can use $\Gamma(\phi\to\eta^\prime\gamma)$,
$\Gamma(\eta^\prime\to\omega\gamma)$ and $\Gamma(\eta^\prime\to\rho\gamma)$
as inputs, along with
$$\Gamma(\eta^\prime\to V\gamma)=\frac{\alpha}{8}\,g^2_{\eta^\prime VV}\,
g^2_{V\gamma}\left (\frac{M_{\eta^\prime}}{f_\pi}\right )^2\left (
1-\frac{M_V^2}{M_{\eta^\prime}^2}\right )^3 M_{\eta^\prime},$$
to obtain
\begin{equation}
g_{\eta^\prime\rho\rho}=0.624\pm 0.044,\;\; 
g_{\eta^\prime\omega\omega}=0.724\pm 0.067,\;\; 
g_{\eta^\prime\phi\phi}=0.886\pm 0.059.
\label{eq2-7}\end{equation}
In terms of these coupling constants, the $F_{\eta^\prime\gamma^*\gamma^*}$ 
form factor has the form
\begin{equation}
F_{\eta^\prime\gamma^*\gamma^*}(k_1^2,k_2^2)=
\frac{g_{\eta^\prime\rho\rho}\,g^2_{\rho\gamma}
M^4_\rho}{(k_1^2-M_\rho^2)(k_2^2-M_\rho^2)}+
\frac{g_{\eta^\prime\omega\omega}\,g^2_{\omega\gamma}
M^4_\omega}{(k_1^2-M_\omega^2)
(k_2^2-M_\omega^2)}+\frac{g_{\eta^\prime\phi\phi}\,g^2_{\phi\gamma}M^4_\phi}
{(k_1^2-M_\phi^2)(k_2^2-M_\phi^2)}.
\label{eq2-8}\end{equation}
Relative phases are again the ones that follow from the nonet-ansatz SU(3)
symmetric interaction Lagrangian
$${\cal{L}}=g\,\epsilon^{\mu\nu\sigma\tau}Sp[(\partial_\mu W_\nu)
(\partial_\sigma W_\tau) P],$$
where
$$W=\frac{1}{\sqrt{2}}\,(\rho^0\lambda_3+\omega_{(8)}\lambda_8)+
\frac{1}{\sqrt{3}}\,\omega_{(1)},\;\; P=\frac{1}{\sqrt{2}}\,
\eta_{(8)}\lambda_8+\frac{1}{\sqrt{3}}\,\eta_{(1)},$$
$\lambda_3,\lambda_8$ being the standard Gell-Mann matrices, and the
$\eta-\eta^\prime$ mixing is given by
$$\eta_{(8)}=\cos{\theta_P}\,\eta+\sin{\theta_P}\,\eta^\prime,\;\;
\eta_{(1)}=\cos{\theta_P}\,\eta^\prime-\sin{\theta_P}\,\eta, 
\;\; \theta_P\approx -20^\circ, $$
while the $\phi-\omega$ mixing is assumed to be ideal:
$$\omega_{(8)}=\sqrt{\frac{2}{3}}\,\phi+\sqrt{\frac{1}{3}}\,\omega,\;\;
\omega_{(1)}=\sqrt{\frac{2}{3}}\,\omega-\sqrt{\frac{1}{3}}\,\phi.$$
Sum rule that follows 
\begin{eqnarray} && 
\Gamma(\eta^\prime\to 2\gamma)=\frac{3M_{\eta^\prime}^3}{2\alpha}\left [ 
\frac{M_{\eta^\prime}}
{M_{\eta^\prime}^2-M_\rho^2}\,\sqrt{\frac{\Gamma(\eta^\prime\to \rho\gamma)\,
\Gamma(\rho\to e^+e^-)M_{\eta^\prime}}{M_\rho({M_{\eta^\prime}^2-M_\rho^2)}}}
\,+\right . \\ && \left . \frac{M_{\eta^\prime}}
{M_{\eta^\prime}^2-M_\omega^2}\,\sqrt{\frac{\Gamma(\eta^\prime\to 
\omega\gamma)\,\Gamma(\omega\to e^+e^-)M_{\eta^\prime}}
{M_\omega({M_{\eta^\prime}^2-M_\omega^2)}}}
+\frac{M_\phi}{M_\phi^2-M_{\eta^\prime}^2}\,\sqrt{\frac{3\Gamma(\phi\to
\eta^\prime\gamma)\,\Gamma(\phi\to e^+e^-)}{M_\phi^2-M_{\eta^\prime}^2}}
\;\right ]^2 \nonumber
\label{eq2-9}\end{eqnarray}
is well satisfied and supports our assumptions. It gives 
$\Gamma(\eta^\prime\to 2\gamma)=(4.56\pm 0.32)~\mathrm{keV}$ 
and the experimental width is \cite{2-25}
$\Gamma(\eta^\prime\to 2\gamma)=(4.29\pm 0.15)~\mathrm{keV}$.

But $\Gamma(P\to 2\gamma)$ widths check the $F_{P\gamma^*\gamma^*}$ form 
factors only for on-shell photons and the two photon contributions into
$e^+e^-\to P\gamma$ we are interested in depend on the off-shell behaviour
of these form factors. Note that (\ref{eq2-5}) and (\ref{eq2-8}) apparently 
violate the QCD prescribed $\sim 1/k^2$ asymptotic behaviour when both 
photon virtualities are large. Therefore now we turn to the $\eta\to\mu^+
\mu^-$ decay, where the off-shell behaviour of the form factor does matter, 
to demonstrate that (\ref{eq2-5}) still gives reasonable estimate.

\section{$\eta\to\mu^+\mu^-$ decay}
There is an intensive literature devoted to the rare decays of pseudoscalar 
mesons into a lepton pair (for a review and references see, for example,
\cite{3-1,3-2,3-2P,3-3,3-4}). Detailed pedagogical calculation in the 
framework of VMD is given in \cite{3-5}. In somewhat different manner, this 
calculation is reproduced in \cite{3-6}. We follow these references in spirit 
but differ in technical details. 

\begin{figure}[htp]
\begin{center}
\begin{picture}(15000,8000)

\drawline\fermion[\W\REG](12000,6000)[6000]
\drawarrow[\E\ATBASE](\pmidx,\pmidy)
\put(\pbackx,\pbacky){\circle*{500}}
\global\advance\pfrontx by 700
\put(\pfrontx,\pfronty){$\mu^-$}
\global\advance\pmidy by 700
\put(\pmidx,\pmidy){$p$}
\put(\pbackx,\pbacky){\circle*{500}}
\drawline\fermion[\S\REG](\pbackx,\pbacky)[6500]
\drawarrow[\N\ATBASE](\pmidx,\pmidy)
\global\advance\pmidx by 700
\put(\pmidx,\pmidy){$p-k$}
\put(\pbackx,\pbacky){\circle*{500}}
\drawline\fermion[\E\REG](\pbackx,\pbacky)[6000]
\drawarrow[\W\ATBASE](\pmidx,\pmidy)
\global\advance\pbackx by 700
\put(\pbackx,\pbacky){$\mu^+$}
\drawline\photon[\NW\REG](\pfrontx,\pfronty)[5]
\put(\pbackx,\pbacky){\circle*{700}}
\drawline\scalar[\W\REG](\photonbackx,\photonbacky)[3]
\global\advance\pbackx by -800
\put(\pbackx,\pbacky){$\eta$}
\global\advance\pmidy by 700
\global\advance\pmidx by -200 
\put(\pmidx,\pmidy){$Q$}
\drawline\photon[\NE\REG](\pfrontx,\pfronty)[5]
\end{picture}
\end{center}
\caption{The leading contribution to the $\eta\to\mu^+\mu^-$ decay.}
\label{fig3-1}\end{figure}
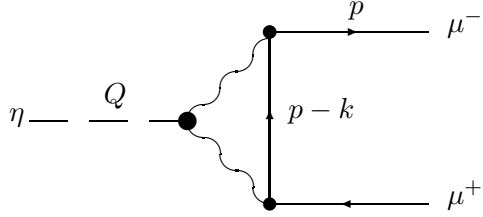

The leading contribution to the $\eta(Q)\to
\mu^+(Q-p,s_+)+\mu^-(p,s_-)$ decay comes from the diagram shown in 
Fig.\ref{fig3-1}.

\noindent The corresponding Feynman amplitude is
\begin{equation}
{\cal{A}}=\frac{e^4}{f_\pi}\,\bar u(p,s_-){\cal{M}}v(Q-p,s_+)=
\left . \left . \frac{e^4}{f_\pi}\,Sp\right [  v(Q-p,s_+)\bar u(p,s_-) 
{\cal{M}}\right ],
\label{eq3-1}\end{equation}
where
$${\cal{M}}=\int\frac{dk}{(2\pi)^4}\,\frac{\gamma_\mu(\hat p-\hat k+m_\mu)
\gamma_\nu\,\epsilon^{\mu\nu\sigma\tau}k_\sigma Q_\tau}{[(p-k)^2-m_\mu^2]
\,k^2\,(Q-k)^2}\,F_{\eta\gamma^*\gamma^*}(k^2,(Q-k)^2). $$
The lepton pair from this decay has total angular momentum $J=0$ and hence
is either in singlet $^1S_0$  or in triplet $^3P_0$ state. But the triplet
state has ${\bf CP}$-parity $(-1)^{s+1}=1$ which does not matches the negative
${\bf CP}$-parity of the $\eta$ meson. Therefore, assuming ${\bf CP}$
invariance, $v(Q-p,s_+)\bar u(p,s_-)$ in (\ref{eq3-1}) can be replaced by
the projection operator to the singlet state for the outgoing $\mu^+\mu^-$
system
$${\cal{P}}(Q-p,p)=\frac{1}{\sqrt{2}}\left [v(Q-p,+)\bar u(p,-)+
v(Q-p,-)\bar u(p,+)\right ].$$  
This projection operator was calculated in \cite{3-2} with the result
\begin{equation}
{\cal{P}}(p_+,p_-)=\frac{1}{2\sqrt{2t}}\left [ -2m_\mu \,(\hat p_++\hat p_-)
\gamma_5+\frac{1}{2}\,\epsilon_{\mu\nu\sigma\tau}(p_-^\sigma p_+^\tau-
p_+^\sigma p_-^\tau)\sigma^{\mu\nu}+t\gamma_5\right ],
\label{eq3-2}\end{equation}
where $t=(p_-+p_+)^2$ and $\sigma^{\mu\nu}=\frac{i}{2}(\gamma^\mu
\gamma^\nu-\gamma^\nu\gamma^\mu)$.
Therefore
$${\cal{A}}=\frac{e^4}{f_\pi}\int\frac{dk}{(2\pi)^4}\,
\frac{\epsilon^{\mu\nu\sigma\tau}k_\sigma Q_\tau L_{\mu\nu}}
{[(p-k)^2-m_\mu^2]\,k^2\,(Q-k)^2}\,F_{\eta\gamma^*\gamma^*}(k^2,(Q-k)^2),$$
with
$$L_{\mu\nu}=Sp\left [{\cal{P}}\gamma_\mu(\hat p-\hat k+m_\mu)\gamma_\nu
\right ]=-2i\sqrt{2}\,\frac{m_\mu}{M_\eta}\,\epsilon_{\mu\nu\sigma\tau}
k^\sigma Q^\tau.$$
But then
$$\epsilon^{\mu\nu\sigma\tau}k_\sigma Q_\tau L_{\mu\nu}=
\frac{8im_\mu}{\sqrt{2}\,M_\eta}\left ( M_\eta^2\, k^2-(k\cdot Q)^2 
\right ).$$
In the above expressions REDUCE \cite{3-7} was used to perform Dirac 
algebra and calculate traces.

Finally, we get in the standard way
\begin{equation}
\Gamma(\eta\to\mu^+\mu^-)=\frac{2\alpha^4}{\pi}M_\eta\left (\frac{m_\mu}
{f_\pi}\right)^2\sqrt{1-\frac{4m_\mu^2}{M_\eta^2}}\,\left |R\right |^2,
\label{eq3-3}\end{equation}
where \cite{3-8,3-9}
\begin{equation}
R=\frac{i}{\pi^2}\int\frac{dk}{M_\eta^2}\,\frac{ M_\eta^2\, k^2-
(k\cdot Q)^2 }{k^2\,(Q-k)^2\,[(p-k)^2-m_\mu^2]}\,
F_{\eta\gamma^*\gamma^*}(k^2,(Q-k)^2).
\label{eq3-4}\end{equation}
Remembering VMD expression (\ref{eq2-5}) for the $F_{\eta\gamma^*\gamma^*}$ 
form factor, we can write
\begin{equation}
R=g_{\eta\rho\rho}\,g^2_{\rho\gamma}\,I(M_\rho^2,M_\rho^2)+
g_{\eta\omega\omega}\,g^2_{\omega\gamma}\,I(M_\omega^2,M_\omega^2)-
g_{\eta\phi\phi}\,g^2_{\phi\gamma}\,I(M_\phi^2,M_\phi^2),
\label{eq3-5}\end{equation}
where the master integral has the form
\begin{equation}
I(M_1^2,M_2^2)=\frac{i}{\pi^2}\int\frac{dk}{M_\eta^2}\,\frac
{ M_\eta^2\, k^2- (k\cdot Q)^2 }{k^2\,(Q-k)^2\,
[(p-k)^2-m_\mu^2]}\,\frac{M_1^2M_2^2}{(k^2-M_1^2)\,[(Q-k)^2-M_2^2]}.
\label{eq3-6}\end{equation}
The imaginary part of this integral can be calculated by using the Cutkosky 
rules \cite{3-10,3-11}. When intermediate masses $M_{1,2}$ are greater than 
the $\eta$-meson mass, only two-photon cut contributes to the discontinuity 
of $I(M_1^2,M_2^2)$ (which is twice its imaginary part) and, therefore,
$$disc\, I(M_1^2,M_2^2)=\frac{(-2\pi i)^2}{\pi^2}\int dk \,
(-\vec{k}^{\,2}\,)\, \frac{\delta_+(k^2)\delta_+[(Q-k)^2]}{k^2-2p\cdot k}\,
\frac{M_1^2M_2^2} {(k^2-M_1^2)\,[(Q-k)^2-M_2^2]},$$
where 
$$\delta_+(k^2)=\Theta(k_0)\delta(k^2)=\frac{1}{2k_0}
\delta(k_0-|\vec{k}|).$$
Because of $\delta$-functions, calculation is straightforward and gives the 
well known \cite{3-1} model-independent result
\begin{equation}
Im\,I(M_1,M_2)=\frac{1}{2}\,disc\, I(M_1^2,M_2^2)=\frac{\pi}{4\beta}\,
\ln{\frac{1-\beta}{1+\beta}},\;\;\beta=\sqrt{1-\frac{4m_\mu^2}{M_\eta^2}}.
\label{eq3-7}\end{equation} 
Calculation of the real part is much more tricky and as a first step involves 
the following algebraic identity
$$
\frac{\left [ k^2Q^2- (k\cdot Q)^2\right ]\, M_1^2\,M_2^2}{D_1D_2D_3D_4D_5}=
-\frac{\lambda(M_1^2,M_2^2,Q^2)}{4D_3D_4D_5}-\frac{(Q^2)^2}
{4D_1D_2D_3}+\frac{(Q^2-M_1^2)^2}{4D_2D_3D_4}+ $$
\begin{equation}
\frac{(Q^2-M_2^2)^2}{4D_1D_3D_5}-
\frac{M_1^2}{4}\,\left [\frac{1}{D_3D_5}-\frac{1}{D_2D_3}\right ]-
\frac{M_2^2}{4}\,\left [\frac{1}{D_3D_4}-\frac{1}{D_1D_3}\right ],
\label{eq3-8}\end{equation}
where
$$D_1=k^2,\;\;D_2=(Q-k)^2,\;\;D_3=k^2-2p\cdot k,\;\;D_4=k^2-M_1^2,\;\;
D_5=(Q-k)^2-M_2^2,$$
and $\lambda(x,y,z)=x^2+y^2+z^2-2xy-2xz-2yz$ is the triangle function. 
Using this identity and introducing the dimensionless variables
$$r=\frac{m_\mu^2}{M_\eta^2},\;\; r_1=\frac{M_1^2}{M_\eta^2},\;\;
r_2=\frac{M_2^2}{M_\eta^2},\;\;\rho_1=\frac{m_\mu^2}{M_1^2},\;\;
\rho_2=\frac{m_\mu^2}{M_1^2}, $$
we get
\begin{equation}
I(M_1^2,M_2^2)=J(r_1,r_2)+J(0,0)-J(r_1,0)-J(0,r_2)-\frac{r_1}{4}
\left [f_1(\rho_2)-f_1(0)\right ]-\frac{r_2}{4}\left [f_2(\rho_1)-
f_2(0)\right ],
\label{eq3-9}\end{equation}
with
$$f_1(\rho_2)=\frac{i}{\pi^2}\int\frac{dk}{[k^2-2p\cdot k]\,
[(Q-k)^2-M_2^2]},\;\; f_2(\rho_1)=\frac{i}{\pi^2}\int\frac{dk}
{[k^2-2p\cdot k]\,[k^2-M_1^2]},$$
and
$$J(r_1,r_2)=\frac{1}{4}\lambda(1,r_1,r_2)\,g(r_1,r_2),\;
g(r_1,r_2)=-\frac{i}{\pi^2}\int\frac{M_\eta^2\,dk}{[k^2-2p\cdot k]\,
[k^2-M_1^2]\,[(Q-k)^2-M_2^2]}.$$
Integrals with two denominators ($f_1$ and $f_2$) are easy to 
calculate by using
$$\frac{1}{AB}=\int\limits_0^1\frac{dx}{[xA+(1-x)B\,]^2}$$
and the dimensionally regularized integral ($\gamma_E$ is the Euler constant)
$$\frac{i}{\pi^2}\int\frac{dk}{[k^2-A]^2}=-\frac{\Gamma\left(\epsilon/2
\right )}{(\pi A)^{\epsilon/2}}\approx -\left (\frac{2}{\epsilon}-
\gamma_E -\ln{\pi}-ln{\,A}\right ),\;\; \epsilon=4-d\to 0.$$
The result is \cite{3-5,3-6}
\begin{equation}
f_1(\rho_2)-f_1(0)=\int\limits_0^1dx\,\ln{\frac{m_\mu^2\,x^2+M_2^2\,(1-x)}
{m_\mu^2\,x^2}}=-\frac{1}{2\rho_2}\left [\ln{\rho_2}+\sqrt{1-4\rho_2}\,
\ln{\frac{1+\sqrt{1-4\rho_2}}{1-\sqrt{1-4\rho_2}}}\,\right ],
\label{eq3-10}\end{equation}
and
\begin{equation}
f_2(\rho_1)-f_2(0)=-\frac{1}{2\rho_1}\left [\ln{\rho_1}+\sqrt{1-4\rho_1}\,
\ln{\frac{1+\sqrt{1-4\rho_1}}{1-\sqrt{1-4\rho_1}}}\,\right ].
\label{eq3-11}\end{equation}
As for the integral with three denominators $g(r_1,r_2)$, we use Feynman
parameterization
$$\frac{1}{D_3D_4D_5}=2\int\limits_0^1 dx \int\limits_0^x \frac{dy}
{[(x-y)D_3+(1-x)D_4+yD_5]^3}$$
and the integral
$$\frac{i}{\pi^2}\int\frac{dk}{[k^2-A]^3}=\frac{1}{2A}$$
to get
$$g(r_1,r_2)=\int\limits_0^1 dx \int\limits_0^x \frac{dy}{-r
\,(x^2+y^2)-\frac{1+\beta^2}{2}\,xy+y+r_1x-r_2y-r_1}.$$
Let us shift the $y$-variable as follows:
$$y\to y+\alpha \,x,\;\; \alpha=-\frac{1-\beta}{1+\beta}.$$ 
Then
$$g(r_1,r_2)=\int\limits_0^1 dx \int\limits_{-\alpha x}^{(1-\alpha)x} 
\frac{dy}{f(y)\,x+h(y)},$$
with
$$f(y)=r_1+\frac{1-\beta}{1+\beta}\,(r_2-1)-\beta y,\;\;
h(y)=-ry^2+(1-r_2)y-r_1.$$
But
$$\int\limits_0^1 dx \int\limits_{-\alpha x}^{(1-\alpha)x} dy=
\int\limits_0^1 dx \int\limits_0^{(1-\alpha)x} dy-
\int\limits_0^1 dx \int\limits_0^{-\alpha x} dy=
\int\limits_0^{1-\alpha} dy \int\limits_{y/(1-\alpha)}^1 dx-
\int\limits_0^{-\alpha} dy \int\limits_{-y/\alpha}^1 dx $$
and after performing the trivial $x$-integration we end up with
$$g(r_1,r_2)=\int\limits_0^{2/(1+\beta)} \frac{dy}{f(y)}\left 
\{\ln{[f(y)+h(y)]}-
\ln{\left [\frac{f(y)\,y\,(1+\beta)}{2}+h(y)\right]}\right \}- $$ 
\begin{equation}\hspace*{13mm}
\int\limits_0^{(1-\beta)/(1+\beta)} \frac{dy}{f(y)}\left \{\ln{[f(y)+h(y)]}-
\ln{\left [\frac{f(y)\,y\,(1+\beta)}{1-\beta}+h(y)\right]}\right \}.
\label{eq3-12}\end{equation}
We need only the real part and, therefore, 
integrals in (\ref{eq3-12}) are of the type ($\alpha,a,b,A,B,C$ are some real 
constants here)
\begin{equation}
F(\alpha;a,b;A,B,C)=Re\,\int\limits_0^\alpha \frac{dy}{ay+b}\ln{(Ay^2+By+C)}.
\label{eq3-13}\end{equation}
A particular case of such type of integral
$$Z=\int\limits_0^1\left . \left . \frac{dy}{y-y_0}\right [\ln{(y-y_1)
(y-y_2)}-\ln{(y_0-y_1)(y_0-y_2)}\right ]$$
was considered in detail in \cite{3-12} (see also \cite{3-13}). Using 
their result
$$Re\,Z=Re\,\left [Li_2\left (\frac{y_0}{y_0-y_1}\right )-
Li_2\left (\frac{y_0-1}{y_0-y_1}\right )+Li_2\left (\frac{y_0}{y_0-y_2}
\right )-Li_2\left (\frac{y_0-1}{y_0-y_2}\right )\right], $$
we can readily calculate (\ref{eq3-13}) in terms of dilogarithms:
$$
F(\alpha;a,b;A,B,C)=\frac{1}{a}\,Re\,\left \{ \ln{\left (1+\alpha\,\frac{a}
{b}\right )}\,\ln{\left (A\,\frac{b^2}{a^2}-B\,\frac{b}{a}+C\right )}+
\right .$$
\begin{equation}
\left . Li_2\left (\frac{b}{b+y_1a}\right )-Li_2\left (\frac{b+\alpha\,a}
{b+y_1a} \right )+Li_2\left (\frac{b}{b+y_2a}\right )-
Li_2\left (\frac{b+\alpha\,a}{b+y_2a}\right )\right \},
\label{eq3-14}\end{equation}
where
$$y_1=\frac{-B+\sqrt{B^2-4AC}}{2A},\;\;y_2=\frac{-B-\sqrt{B^2-4AC}}{2A}.$$
Returning to (\ref{eq3-12}), we get finally
\begin{eqnarray}
Re\,g(r_1,r_2)=&&F\left (\frac{2}{1+\beta};-\beta,r_1+\frac{1-\beta}
{1+\beta}\,(r_2-1);-r,1-\beta-r_2,\frac{1-\beta}{1+
\beta}\,(r_2-1) \right )- \nonumber \\ &&
F\left (\frac{1-\beta}{1+\beta};-\beta,r_1+\frac{1-\beta}
{1+\beta}\,(r_2-1);-r,1-\beta-r_2,\frac{1-\beta}{1+
\beta}\,(r_2-1) \right )- \nonumber \\ &&
F\left (\frac{2}{1+\beta};-\beta,r_1+\frac{1-\beta}
{1+\beta}\,(r_2-1);-\frac{(1+\beta)^2}{4},\frac{1+\beta}{2}(1+r_1-r_2),
-r_1 \right )+ \nonumber \\ &&
F\left (\frac{1-\beta}{1+\beta};-\beta,r_1+\frac{1-\beta}
{1+\beta}\,(r_2-1);-\frac{(1+\beta)^3}{4(1-\beta)},\frac{1+\beta}{1-\beta}
\,r_1,-r_1 \right ).
\label{eq3-15}\end{eqnarray}
Some arguments of dilogarithms in (\ref{eq3-15}) are complex. Therefore for
numerical evaluation we need an algorithm to evaluate this function for
complex argument. We use the algorithm described in \cite{3-12} (see 
\cite{3-14} for another algorithm). 

First of all the argument of the dilogarithm is brought in the region 
$|x|\le1,\; -1\le Re(x)\le\frac{1}{2}$ by using the functional identities
$$Li_2(x)=-Li_2(1-x)+\frac{\pi^2}{6}-\ln{(x)}\,\ln{(1-x)}, \;\;
Li_2(x)=-Li_2\left (\frac{1}{x}\right )-\frac{\pi^2}{6}-\frac{1}{2}
\ln^2{(-x)}.$$
Then the series expansion
$$Li_2(x)=\sum\limits_{n=0}^\infty B_n\,\frac{z^{n+1}}{(n+1)!},$$
truncated at $n=20$, is used with $B_n$ Bernoulli numbers and 
$z=-\ln{(1-x)}$.

Now we have all ingredients at hand to calculate numerically 
$\Gamma(\eta\to\mu^+\mu^-)$. The result is
\begin{equation}
Br(\eta\to\mu^+\mu^-)=\frac{\Gamma(\eta\to\mu^+\mu^-)}{\Gamma_\eta}=
(5.2\pm 1.2)\times 10^{-6}.
\label{eq3-16}\end{equation}
The quoted uncertainty is dominated by uncertainty in the $g_{\eta\rho\rho}$
coefficient. The experimental number is
$Br(\eta\to\mu^+\mu^-)=(5.8\pm 0.8)\times 10^{-6}$
and again we observe a good agreement. However, it should be mentioned that
the dominant contribution into (\ref{eq3-16}) is given by the imaginary part 
of the amplitude which is in fact model independent:
$$Re(R)\approx -0.015,\;\;\; Im(R)\approx -0.074.$$
To check our formulas and computer code, we have make sure that they reproduce 
numerical values of $2I(M_\rho^2,M_\rho^2)$ and of the analogous integral 
for the $\pi^0\to e^+e^-$ decay given in \cite{3-1}.

Analogous considerations apply to the $\eta^\prime\to\mu^+\mu^-$ decay if we
use the form factor (\ref{eq2-8}) with coupling constants (\ref{eq2-7}). But
now the imaginary part of the amplitude has an additional contribution from
the on-shell $\rho\gamma$ and $\omega\gamma$ intermediate states. This
contribution is calculated by using the Cutkosky rules with the result
\begin{equation}
\Delta Im\, I(M_1^2,M_2^2)=-\frac{\pi}{4\beta}\left [ \left ( 1-
\frac{M_1^2}{M_{\eta^\prime}^2} \right )^2+\left ( 1-\frac{M_2^2}
{M_{\eta^\prime}^2} \right )^2\right ]\ln{\frac{1-\beta}{1+\beta}}.
\label{eq3-17}\end{equation}
The numerical calculation yields the branching ratio
\begin{equation}
Br(\eta^\prime\to\mu^+\mu^-)=(1.4\pm 0.2)\times 10^{-7}.
\label{eq3-18}\end{equation}
Now the imaginary part of the amplitude is only $1.5$-times larger in 
magnitude than the real part:
$$Re(R)\approx 0.064,\;\;\; Im(R)\approx -0.093.$$
Unfortunately there is no experimental number for the $\eta^\prime\to\mu^+
\mu^-$ branching ratio to compare with (\ref{eq3-18}). 

\section{Two-photon contribution to $e^+e^-\to\eta\gamma$: constant 
form-factor}
The amplitude of the $e^+(p_+,s_+)+e^-(p_-,s_-)\to \eta(q)+
\gamma(k,\epsilon)$ process is given by
$${\cal A}=\bar v (p_+,s_+)\Gamma^\mu(p_+,p_-,k)u(p_-,s_-)\,\epsilon^*_\mu,$$
where gauge symmetry and invariance under parity and charge conjugation
dictate the following decomposition of $\Gamma^\mu$ in terms of four 
independent invariant form factors (up to irrelevant additional structures 
proportional to $k^\mu$) \cite{3-15}
$$\Gamma^\mu=P(\chi_-,\chi_+)\,\left [(k\cdot p_+)\,p_-^\mu-(k\cdot p_-)
\,p_+^\mu \right ]\gamma_5-i\,T(\chi_-,\chi_+)\,\sigma^{\mu\nu}
k_\nu\gamma_5+$$
\begin{equation}
A_+(\chi_-,\chi_+)\,\left [\hat k \,p_+^\mu-(k\cdot p_+)\gamma^\mu 
\right ]\gamma_5- A_-(\chi_-,\chi_+)\,\left [\hat k \,p_-^\mu-(k\cdot p_-)
\gamma^\mu \right ]\gamma_5,
\label{eq3-19}\end{equation}
with
\begin{equation}
P(\chi_-,\chi_+)=P(\chi_+,\chi_-),\;\;A_\pm(\chi_-,\chi_+)=
A_\mp(\chi_+,\chi_-),\;\; T(\chi_-,\chi_+)=T(\chi_+,\chi_-).
\label{eq3-20}\end{equation}
Here $\chi_-=2\,k \cdot p_-$ and $\chi_+=2\,k \cdot p_+$.

The invariant form factors $P,A_\pm$ and $T$ can be projected out from 
$\Gamma^\mu$ by a suitable projector operator $\Lambda^\mu_F,\; F=P,A_\pm,T$
according to the the formula
\begin{equation}
F=Sp\left [\Lambda^\mu_F\, (\hat p_+-m_e)\,\Gamma_\mu\, 
(\hat p_-+m_e) \right ] 
\label{eq3-21}\end{equation}
We are interested in the $m_e\to 0$ limit and in this limit these projectors
have the form (our expressions differ somewhat from ones given in 
\cite{3-15})  
$$\Lambda^\mu_P=\frac{1}{2\Delta^2\delta^2}\left [
-i\,\epsilon^{\mu\nu\sigma\tau}k_\nu\,p_{+\sigma}\,p_{-\tau}
-\left ( \frac{\Delta^2}{\delta^2\,k\cdot p_+\,k\cdot p_-}-2 \right ) 
\Delta^\mu\gamma_5\right ],$$
$$\hspace*{-5mm}
\Lambda^\mu_{A_\pm}=\left .\frac{\mp 1}{16(\Delta^2)^2}\right \{\left [ 
2\Delta^2\pm p^2\,k\cdot (p\mp \delta)\,k\cdot \delta\right ]\left [
k\cdot p\, \gamma^\mu-\hat k\,p^\mu\right ]\pm$$ $$\left . p^2\left (
2\,\frac{\Delta^2}{\delta^2}-k\cdot (p\mp \delta)\,k\cdot p\right )\left (
k\cdot\delta\,\gamma^\mu-\hat k\,\delta^\mu \right )\right \}\gamma_5,$$
\begin{equation} \hspace*{-70mm}
\Lambda^\mu_T=\frac{i}{4\Delta^2}\,\epsilon^{\mu\nu\sigma\tau}k_\nu\, 
p_{+\sigma}\,p_{-\tau},
\label{eq3-22}\end{equation}
where
$$\Delta=(k\cdot p_+)\,p_--(k\cdot p_-)\,p_+,\;\;\delta=p_+-p_-,\;\;
p=p_++p_-.$$
Using these projector operators and calculating traces by means of REDUCE
\cite{3-7}, it can be found that $P$ and $T$ form factors vanish in the 
$m_e\to 0$ limit for diagrams shown in Fig.\ref{fig1-1}. Therefore, in 
this limit we get in the standard way
\begin{equation}
\frac{d\sigma^{(2\gamma)}\,(e^+e^-\to\eta\gamma)}{d\Omega}=\frac{s^2}
{4^5\,\pi^2}\left (1-\frac{M_\eta^2}{s}\right )^3\left [ (1+\cos{\theta})^2
\,|A_+|^2+(1-\cos{\theta})^2\,|A_-|^2\right ],
\label{eq3-23}\end{equation}
where $\sigma^{(2\gamma)}\,(e^+e^-\to\eta\gamma)$ denotes purely two-photon 
contribution into the $e^+e^-\to\eta\gamma$ cross section, $s=(p_-+p_+)^2$
and $\theta$ is the flight angle of the photon with respect to the electron 
momentum $\vec{p}_-$ in the center of mass frame.

To calculate (\ref{eq3-23}), we need to find out only the $A_-(\chi_-,
\chi_+)$ form factor, because of the charge conjugation symmetry relations
(\ref{eq3-20}). The following method \cite{1-3} for the form factor 
determination seems to be more convenient than the direct use of the suitable 
projector.

In the massless electron limit
\begin{equation}
\Gamma^\mu=[A_1\,p_-^\mu-A_2\,p_+^\mu]\,\hat k\,\gamma_5+
A_3\,\gamma^\mu\gamma_5,
\label{eq3-24}\end{equation}
where $A_1=-A_-,\;A_2=-A_+$ and $A_3=k\cdot p_+\,A_2-k\cdot p_+\,A_2$.
Using the identity
$$\gamma_\mu\,\epsilon^{\mu\nu\sigma\tau}=\frac{1}{2}\left (\gamma^\nu
\gamma^\sigma\gamma^\tau-\gamma^\tau\gamma^\sigma\gamma^\nu\right )
(i\gamma_5)$$
it can be shown \cite{1-3} that the first two diagrams from Fig.\ref{fig1-1}
contribute to $A_3$ alone. Therefore to determine $A_-=-A_1$ it is sufficient
to consider only the third diagram with photon emission from the internal
line. The contribution of this diagram in the massless electron limit looks 
like
\begin{equation}
\Gamma_\mu^{(c)}=\frac{e^5}{f_\pi}\int\frac{dl}{(2\pi)^4}\,\frac{\gamma_\nu
(\hat p_+-\hat l)\gamma_\mu(\hat p_+-\hat l-\hat k)\gamma_\lambda\,
\epsilon^{\nu\lambda\sigma\tau}l_\sigma q_\tau}{l^2\,(q-l)^2\,(p_+-l)^2\,
(p_+-l-k)^2}\,F_{\eta\gamma^*\gamma^*}(l^2,(q-l)^2).
\label{eq3-25}\end{equation}
As follows from (\ref{eq3-24}), the coefficient of the 
$k_\nu p_{-\mu}$ term in
$Sp(\gamma_\nu\hat p_-\Gamma_\mu\gamma_5\,\hat p_+)$ is $-4\,p_-\cdot p_+\,
A_1=4\,p_-\cdot p_+\,A_-$. Besides, (\ref{eq3-25}) is free from ultraviolet
divergences even for constant form factor $F_{\eta\gamma^*\gamma^*}$ (
point-like $\eta$). Therefore in this section we assume, as in \cite{1-3}, 
that the $F_{\eta\gamma^*\gamma^*}$ is just a constant.
Then, to find out the $A_-$ form factor, we 
combine $D_1=l^2$, $D_2=(q-l)^2$, $D_3=(p_+-l)^2$ and $D_4=(p_+-l-k)^2$ 
denominators in (\ref{eq3-25}) by using 
$$\frac{1}{D_1\,D_2\,D_3\,D_4}=3!\int\limits_0^1 dx_1\cdots
\int\limits_0^1 dx_4\,\delta\left ( \sum x_i -1\right )\frac{1}
{[x_1D_1+x_2D_2+x_3D_3+x_4D_4]^4},$$
shift the variables in the momentum integral according to
$$l\to l+x_2p_-+(1-x_1)p_+-(x_2+x_4)k,$$
calculate the trace $Sp(\gamma_\nu\hat p_-\Gamma^{(c)}_\mu\gamma_5\,
\hat p_+)$ with the help of REDUCE \cite{3-7}, replace in resulting momentum 
integrals
$$l_\mu l_\nu\to\frac{1}{4}g_{\mu\nu}\,l^2,\;\; l_\mu l_\nu l_\lambda \to 0,
\;\; l_\mu\to 0,$$
and separate the coefficient of the $k_\nu p_{-\mu}$ term. As a result we 
obtain 
$$A_-=\frac{6e^5F_{\eta\gamma^*\gamma^*}}{f_\pi}\int\limits_0^1 dx_1
\cdots\int\limits_0^1 dx_4\,\delta\left ( \sum x_i -1\right ) \times $$
\begin{equation}
\frac{i}{(2\pi)^4}\int dl\,\frac
{(1-4x_2-2x_4)\,l^2-2sx_2^2x_3+2\chi_+(x_2+x_4-1)(x_2x_3-x_1x_4)}
{[l^2-x_2x_3\chi_--x_1x_4\chi_++M_\eta^2\,x_1x_2]^4},
\label{eq3-26}\end{equation}
Our result (\ref{eq3-26}) differs somewhat from the corresponding expression 
in \cite{1-3}. For this reason we have cross-checked it by an independent 
calculation where part of the above given program was performed manually. 

Momentum integrals read
$$\frac{i}{\pi^2}\int dl\,\frac{l^2}{[l^2-A]^4}=\frac{1}{3A},\;\;
\;\;\frac{i}{\pi^2}\int dl\,\frac{1}{[l^2-A]^4}=-\frac{1}{6A^2}.$$
Therefore
$$A_-=\frac{e^5F_{\eta\gamma^*\gamma^*}}{8\pi^2f_\pi}\int\limits_0^1 dx_1
\cdots\int\limits_0^1 dx_4\, \delta\left ( \sum x_i -1\right )\times $$ 
$$\left [\frac{1-4x_2-2x_4}{A}+\frac{s\,x_2^2x_3-\chi_+(x_2+x_4-1)(x_2x_3-
x_1x_4)}{A^2}\right ],$$
where
$$A=x_2x_3\chi_-+x_1x_4\chi_+-M_\eta^2x_1x_2.$$
Let us introduce dimensionless variables 
\begin{equation}
X_+=1+\frac{\chi_+}{M_\eta^2},\;\;\;\;X_-=1+\frac{\chi_-}{M_\eta^2},
\label{eq3-27}\end{equation}
and note that
$$\frac{1}{M_\eta^2}\,\frac{\partial}{\partial X_+}\,\frac{1}{A}=
-\frac{x_1x_4}{A^2},\;\;\;\; \frac{1}{M_\eta^2}\,\frac{\partial}
{\partial X_-}\,\frac{1}{A}=-\frac{x_2x_3}{A^2}.$$
Then
\begin{equation}
A_-=A_-^{(1)}-A_-^{(2)},
\label{eq3-28}\end{equation}
with
$$M_\eta^2A_-^{(1)}=\frac{e^5F_{\eta\gamma^*\gamma^*}}{8\pi^2f_\pi}\left \{
\left [ 2+(X_++X_--1)\frac{\partial}{\partial X_-}\right ] I_2(X_+,X_-)+
\right .$$ $$ \left .
\left [ 2+(1-X_+)\left (\frac{\partial}{\partial X_-}-
\frac{\partial}{\partial X_+}\right )\right ][I_2(X_+,X_-)+I_4(X_+,X_-)]
\right \},$$
and
$$M_\eta^2A_-^{(2)}=\frac{e^5F_{\eta\gamma^*\gamma^*}}{8\pi^2f_\pi}\left [
1+(1-X_+)\left (\frac{\partial}{\partial X_-}-\frac{\partial}{\partial X_+}
\right )\right ]I_0(X_+,X_-). $$
Here the master integrals are
$$\hspace*{-7mm}
I_0(X_+,X_-)=\int\limits_0^1 dx_1 \cdots\int\limits_0^1 dx_4\, \delta
\left ( \sum x_i -1\right )\frac{1}{x_1x_2+(1-X_-)x_2x_3+(1-X_+)x_1x_4},$$
\begin{equation}
I_n(X_+,X_-)=\int\limits_0^1 dx_1 \cdots\int\limits_0^1 dx_4\, \delta
\left ( \sum x_i -1\right )\frac{x_n}{x_1x_2+(1-X_-)x_2x_3+(1-X_+)x_1x_4}.
\label{eq3-29}\end{equation}
In (\ref{eq3-28}) the $A_-^{(1)}$ part is the same as in \cite{1-3}, while 
the $A_-^{(2)}$ part is absent in \cite{1-3}.

The master integrals (\ref{eq3-29}) can be considered as integrals over a
tetrahedron shaped 3-dimensional domain in the $(x_1,x_2,x_3)$-space. By 
means of variable transformation \cite{1-3}
\begin{equation}
x_1=xz,\;\; x_2=y(1-z),\;\; x_3=z(1-x),\;\; x_4=(1-z)(1-y),
\label{eq3-30}\end{equation}
with the Jacobian
$$\left | \frac{\partial(x_1,x_2,x_3)}{\partial(x,y,z)}\right |=z(1-z),$$
the integration domain transforms into a unit cube in the
$(x,y,z)$-space and we obtain
$$I_2=\int\limits_0^1 dx\int\limits_0^1 dy \int\limits_0^1 dz\,\frac{y(1-z)}
{(1-X_+)x+(1-X_-)y+(X_-+X_+-1)xy}=$$
\begin{equation}
\frac{1}{2}\int\limits_0^1 dy\, 
\frac{\ln{[1-X_++X_+y+i\epsilon]}-\ln{[(1-X_-)y+i\epsilon]}}
{1-X_++(X_-+X_+-1)y}\, y.
\label{eq3-31}\end{equation}
Analogously
\begin{equation}
I_2+I_4=\frac{1}{2}I_0=\frac{1}{2}\int\limits_0^1 dx\, 
\frac{\ln{[1-X_-+X_-x+i\epsilon]}-\ln{[(1-X_+)x+i\epsilon]}}
{1-X_-+(X_-+X_+-1)x}.
\label{eq3-32}\end{equation}
In the above formulas the $i\epsilon$ prescription for propagators was 
explicitly restored. 

First of all let us calculate the imaginary part of the amplitude which is
in fact model independent as far as it is dominated by the on-shell 
two-photon intermediate state. In (\ref{eq3-31}) and(\ref{eq3-32}) the 
imaginary parts originate from negative arguments of the logarithms. 
Therefore
$$Im\,(I_2+I_4)=-\frac{\pi}{2}\int\limits_{1-1/X_-}^1\frac{dx}
{1-X_-+(X_-+X_+-1)x}=$$ $$-\frac{\pi}{2(X_-+X_+-1)}\ln{\frac{X_-X_+}
{(X_--1)(X_+-1)}},$$ 
and
$$\hspace*{-10mm} Im\,I_2=-\frac{\pi}{2}\int\limits_{1-1/X_+}^1\frac{y\,dy}
{1-X_++(X_-+X_+-1)y}=$$ $$-\frac{\pi}{2X_+}\left [\frac{1}{X_++X_--1}+
\frac{X_+(X_+-1)}{(X_++X_--1)^2}\,\ln{\frac{X_-X_+}{(X_--1)(X_+-1)}}
\right ].$$  
Inserting these expressions into (\ref{eq3-28}) we get finally
$$
Im\,A_-=\frac{e^5F_{\eta\gamma^*\gamma^*}}{8\pi f_\pi M_\eta^2}\,\frac{
X_+-X_-}{X_-X_+(X_--1)}=$$
\begin{equation}
\frac{e^5F_{\eta\gamma^*\gamma^*}}{\pi f_\pi}\,\frac{M_\eta^2}{s^2}\,
\frac{1}{1-\cos{\theta}}\,\frac{\cos{\theta}}{\left (1+\frac{M_\eta^2}
{s}\right )^2-\left (1-\frac{M_\eta^2}{s}\right )^2\cos^2{\theta}},
\label{eq3-33}\end{equation}
and the corresponding contribution into $\sigma^{(2\gamma)}\,(e^+e^-
\to\eta\gamma)$:
\begin{equation}
\frac{d\sigma^{(2\gamma)}_I}{d\Omega}=\frac{
8\alpha^3}{M_\eta^2}\,\left (\frac{M_\eta^2}{s}\right )^2\,\left (
1-\frac{M_\eta^2}{s}\right )^3\,\frac{\Gamma(\eta\to\gamma\gamma)}
{M_\eta}\,\frac{\cos^2{\theta}}{\left [\left (1+\frac{M_\eta^2}{s}
\right )^2-\left (1-\frac{M_\eta^2}{s}\right )^2\cos^2{\theta}\right ]^2},
\label{eq3-34}\end{equation}
where the constant $F_{\eta\gamma^*\gamma^*}$ was expressed trough the 
two-photon width $\Gamma(\eta\to\gamma\gamma)$. Integrating in the limits
$30^\circ\le\theta\le150^\circ$, and assuming $s=112~\mathrm{GeV}^2$ we 
find numerically 
\begin{equation}
\sigma^{(2\gamma)}_I(e^+e^-\to\eta\gamma)\approx 3.4\cdot 10^{-4}~
\mathrm{fb},\;\;\;\;\sigma^{(2\gamma)}_I(e^+e^-\to\eta^\prime\gamma)\approx
4.5\cdot 10^{-3}~\mathrm{fb}.
\label{eq3-35}\end{equation}
These numbers are quite small compared to the reported BABAR cross sections 
\cite{1-1} 
$$\sigma(e^+e^-\to\eta\gamma)=4.5^{+1.2}_{-1.1}\pm0.3~\mathrm{fb},
\;\;\;\;\sigma(e^+e^-\to\eta^\prime\gamma)=5.4\pm0.8\pm0.3~\mathrm{fb}.$$
But the amplitude has also a real part which in fact dominates for the 
constant $F_{\eta\gamma^*\gamma^*}$ form factor.

The real part of the integral (\ref{eq3-32}) can be evaluated by using
a substitution $t=1-X_-+(X_-+X_+-1)\,x$ and the integral
$$\int\limits_A^B\frac{\ln{(1-Ct)}}{t}\,dt=Li_2(CA)-Li_2(CB),$$
$A$,$B$ and $C$ being some constants. As a result, we get (note that
$Li_2(1)=\pi^2/6$) 
$$Re\,(I_2+I_4)=Re\,\frac{Li_2\left (\frac{X_-}{1-X_+}\right )+
Li_2\left (\frac{X_+}{1-X_-}\right )-Li_2\left (\frac{X_-X_+}
{(1-X_+)(1-X_-)}\right )-\frac{\pi^2}{6}}{2(X_-+X_+-1)}.$$
By using the Abel identity \cite{3-16,3-17} (still valid for real parts if 
$x>1$ and $y>1$)
$$\ln{(1-x)}\,\ln{(1-y)}=Li_2\left (\frac{x}{1-y}\right )+
Li_2\left (\frac{y}{1-x}\right )-Li_2(x)-Li_2(y)-Li_2\left (
\frac{xy}{(1-x)(1-y)}\right ),$$
the above expression can be rewritten as
\begin{equation}
Re\,(I_2+I_4)=\frac{Re\,G(X_-,X_+)}{2(X_-+X_+-1)},
\label{eq3-36}\end{equation}
where
\begin{equation}
G(X_-,X_+)=\ln{(1-X_+)}\,\ln{(1-X_-)}+Li_2(X_+)+Li_2(X_-)
-\frac{\pi^2}{6}. 
\label{eq3-G}\end{equation}
The real part of the integral (\ref{eq3-31}) can be calculated analogously 
with the result
\begin{equation}
Re\,I_2=\frac{1}{2(X_-+X_+-1)}\,Re\,\left [ \ln{\frac{X_+-1}{X_--1}}-
\frac{\ln{(X_+-1)}}{X_+}+\frac{(X_+-1)\,G(X_-,X_+)}{X_-+X_+-1}\right ].
\label{eq3-37}\end{equation}
Using (\ref{eq3-36}) and (\ref{eq3-37}), we find
$$Re\,A_-^{(1)}=\frac{e^5F_{\eta\gamma^*\gamma^*}}{8\pi^2f_\pi}\left \{
\frac{G(X_-,X_+)}{s}-\frac{1}{4k\cdot p_-}\right \},$$
in agreement with \cite{1-3}. But according to our results $A_-^{(1)}$ is 
only a part of the $A_-$ form factor, for which we obtain
\begin{equation}  
Re\,A_-=\frac{e^5F_{\eta\gamma^*\gamma^*}}{8\pi^2f_\pi}\,\frac{1}{M_\eta^2
(X_--1)}\,\left [\ln{\frac{X_+-1}{X_--1}}+\frac{\ln{(X_--1)}}{X_-}-
\frac{\ln{(X_+-1)}}{X_+}-\frac{1}{2}\right ].
\label{eq3-38}\end{equation}
In contrast to \cite{1-3}, all dilogarithms have canceled out.

Having at hand $Re\,A_-$, we can find the corresponding contribution
in the $\sigma^{(2\gamma)}\,(e^+e^-\to\eta\gamma)$ cross section:
\begin{equation}
\frac{d\sigma^{(2\gamma)}_R}{d\Omega}=\frac{\alpha^3}{4\pi^2M_\eta^2}\,
\left (1-\frac{M_\eta^2}{s}\right )\,\frac{\Gamma(\eta\to\gamma\gamma)}
{M_\eta}\,\left [\, |F(\cos{\theta})|^2+|F(-\cos{\theta})|^2\right ],
\label{eq3-39}\end{equation}
where
$$F(\cos{\theta})=\ln{\frac{1+\cos{\theta}}{1-\cos{\theta}}}
+\frac{\ln{(X_--1)}}{X_-}-\frac{\ln{(X_+-1)}}{X_+}-\frac{1}{2}$$
and
\begin{equation}
X_\pm=1+\frac{s}{2M_\eta^2}\left (1-\frac{M_\eta^2}{s}\right )(1\pm
\cos{\theta}).
\label{eq3-40}\end{equation}
Numerically, for $30^\circ\le\theta\le150^\circ$ and $s=112~\mathrm{GeV}^2$, 
we find
\begin{equation}
\sigma^{(2\gamma)}_R(e^+e^-\to\eta\gamma)\approx 0.5~\mathrm{fb},\;\;\;\;
\sigma^{(2\gamma)}_R(e^+e^-\to\eta^\prime\gamma)\approx 0.8~\mathrm{fb}.
\label{eq3-41}\end{equation}
These cross sections are quite substantial and they could cause problems 
with the BABAR analysis if the constant $F_{\eta\gamma^*\gamma^*}$ form 
factor would not be unphysical. In reality, however, the form factor drops
quickly at large photon virtualities. Therefore in the next section we
consider more realistic VMD form factor.
 
\section{Two-photon contribution to $e^+e^-\to\eta\gamma$: Vector Meson
Dominance model}
The VMD $F_{\eta\gamma^*\gamma^*}$ form factor is given by (\ref{eq2-5}).
Its insertion into (\ref{eq3-25}) produces momentum integrals with six 
denominators. Their treatment becomes less formidable if we use the identity
$$\frac{M^4}{D_1D_2D_3D_4D_5D_6}=\frac{1}{D_1D_2D_3D_4}+\frac{1}{D_3D_4D_5
D_6}-\frac{1}{D_1D_3D_4D_6}-\frac{1}{D_2D_3D_4D_5},$$
with
$$D_1=l^2,\;D_2=(q-l)^2,\;D_3=(p_+-l)^2,\;D_4=(p_+-l-k)^2,\;
D_5=l^2-M^2,\;D_6=(q-l)^2-M^2.$$
Then the procedure described in the previous section gives
\begin{equation}
A_-=\frac{e^5}{8\pi^2f_\pi}\left [\,g_{\eta\rho\rho}\,g^2_{\rho\gamma}\,
\tilde I(M_\rho^2,M_\rho^2)+g_{\eta\omega\omega}\,g^2_{\omega\gamma}\,
\tilde I(M_\omega^2,M_\omega^2)-g_{\eta\phi\phi}\,g^2_{\phi\gamma}\,
\tilde I(M_\phi^2,M_\phi^2)\,\right ],
\label{eq4-1}\end{equation}
with
\begin{equation}
\tilde I(M_1^2,M_2^2)=\tilde J(M_1^2,M_2^2)+\tilde J(0,0)-
\tilde J(M_1^2,0)-\tilde J(0,M_2^2).
\label{eq4-2}\end{equation}
Here
$$M_\eta^2\,\tilde J(M_1^2,M_2^2)=\left [ 2+(X_++X_--1)\frac{\partial}
{\partial X_-}\right ] \tilde I_2(X_+,X_-)+$$
$$\left [ 2+(1-X_+)\left (\frac{\partial}
{\partial X_-}-\frac{\partial}{\partial X_+}\right )\right ]
[\,\tilde I_2(X_+,X_-)+\tilde I_4(X_+,X_-)\,]-$$
\begin{equation}
\left [ 1+(1-X_+)\left (\frac{\partial}{\partial X_-}-
\frac{\partial}{\partial X_+}\right )\right ]\tilde I_0(X_+,X_-),
\label{eq4-3}\end{equation}
and the new master integrals are
$$\hspace*{-7mm}
\tilde I_0(X_+,X_-)=\int\limits_0^1 dx_1 \cdots\int\limits_0^1 dx_4\, \frac{
\delta \left ( \sum x_i -1\right )}{x_1x_2+(1-X_-)x_2x_3+(1-X_+)x_1x_4
-x_1r_1-x_2r_2},$$
\begin{equation}
\tilde I_n(X_+,X_-)=\int\limits_0^1 dx_1 \cdots\int\limits_0^1 dx_4\, \frac{
\delta \left ( \sum x_i -1\right )\, x_n}{x_1x_2+(1-X_-)x_2x_3+(1-X_+)x_1x_4
-x_1r_1-x_2r_2}.
\label{eq4-4}\end{equation}
The presence of the additional $x_1r_1+x_2r_2$ term in the denominators 
compared to (\ref{eq3-29}) makes the evaluation of these integrals, 
of course, more complicated but not substantially different from the 
evaluation of (\ref{eq3-29}). Therefore we present only final results. 
$$ \hspace*{-58mm} Re\,(\tilde I_2+\tilde I_4)=\frac{1}{(X_-+X_+-1)}\,Re\,
\int\limits_0^1 dz\,(1-z)\,G(X^\prime_-,X^\prime_+),$$
$$Re\,\tilde I_2=\frac{1}{(X_-+X_+-1)}Re\int\limits_0^1 dz\,(1-z)\,
\left [ \ln{\frac{X^\prime_+-1}{X^\prime_--1}}-
\frac{\ln{(X^\prime_+-1)}}{X^\prime_+}+\frac{(X^\prime_+-1)\,
G(X^\prime_-,X^\prime_+)}{X^\prime_-+X^\prime_+-1}\right ], $$
\begin{equation}
\hspace*{-85mm} Re\,\tilde I_0=\frac{1}{(X_-+X_+-1)}\,Re\,
\int\limits_0^1 dz\,G(X^\prime_-,X^\prime_+).
\label{eq4-5}\end{equation} 
Here the function $G(x,y)$ is defined by (\ref{eq3-G}) and 
\begin{equation}
X^\prime_-=\frac{z\,[r_1-(1-z)X_-]}{r_1z+r_2\,(1-z)-z(1-z)},\;\;\;\;
X^\prime_+=\frac{(1-z)\,[r_2-zX_+]}{r_1z+r_2\,(1-z)-z(1-z)}.
\label{eq4-6}\end{equation}  
We need only the real parts because the imaginary part of the amplitude
is dominated by the two-photon cut and, therefore, is the same as for the 
constant $F_{\eta\gamma^*\gamma^*}$ form factor.

Using (\ref{eq4-5}), we get after some algebra
$$\hspace*{-22mm}M_\eta^2Re\,\tilde J(M_1^2,M_2^2)=\frac{M_\eta^2}{s}\,
Re\,\int\limits_0^1 dz (1-2z)\,G(X_-^\prime,X_+^\prime)+$$
\begin{equation}
Re\,\int\limits_0^1 dz\, \frac{z(1-z)}{r2+z(X_--1)}\left [ \frac{r_1+X_+-1}
{r_1+(X_+-1)(1-z)}\,L(X_-^\prime,X_+^\prime)-1\right ],
\label{eq4-7}\end{equation}
where
$$L(x,y)=\ln{\frac{y-1}{x-1}}+\frac{\ln{(x-1)}}{x}-\frac{\ln{(y-1)}}{y}.$$
Note that
$$M_\eta^2\,\left [r_2+z(X_--1)\right ]=\frac{s}{2}\left [\frac{2M_2^2}
{s}+z\left (1-\frac{M_\eta^2}{s}\right )(1-\cos{\theta})\right],$$
and therefore
$$Re\,\tilde J(M_1^2,M_2^2)=\frac{2}{s}\,Re\,j(r_1,r_2),$$
where the dimensionless function $j(r_1,r_2)$ is given by
$$
\hspace*{-62mm} j(r_1,r_2)=\frac{1}{2}\int\limits_0^1 dz (1-2z)\,
G(X_-^\prime,X_+^\prime)+$$
\begin{equation}
\int\limits_0^1 dz\, \frac{z(1-z)}{\frac{2M_2^2}{s}+z\left (1-
\frac{M_\eta^2}{s}\right )(1-\cos{\theta})}\left [ \frac{r_1+X_+-1}
{r_1+(X_+-1)(1-z)}\,L(X_-^\prime,X_+^\prime)-1\right ].
\label{eq4-8}\end{equation}
The corresponding contribution in the $\sigma^{(2\gamma)}\,(e^+e^-\to\eta
\gamma)$ cross section looks like
\begin{equation}
\frac{d\sigma^{(2\gamma)}_R(e^+e^-\to\eta\gamma)}{d\Omega}=
\frac{\alpha^5}{16\pi f_\pi^2}\,\left (1-\frac{M_\eta^2}{s}\right )^3
\left [\, |f(\cos{\theta})|^2+|f(-\cos{\theta})|^2\right ],
\label{eq4-9}\end{equation}
where
$$f(\cos{\theta})=(1-\cos{\theta})\left [\,g_{\eta\rho\rho}\,
g^2_{\rho\gamma}\,i(M_\rho^2,M_\rho^2)+g_{\eta\omega\omega}\,
g^2_{\omega\gamma}\,i(M_\omega^2,M_\omega^2)-
g_{\eta\phi\phi}\,g^2_{\phi\gamma}\,i(M_\phi^2,M_\phi^2)\,\right ],$$
and
$$i(M_1^2,M_2^2)=j(r_1,r_2)+j(0,0)-j(r_1,0)-j(0,r_2).$$
Now we have all ingredients to find the cross section under interest 
numerically. The results are (for the same kinematic conditions as before)
\begin{equation}
\sigma^{(2\gamma)}_R(e^+e^-\to\eta\gamma)\approx 1.6\cdot 10^{-3}~
\mathrm{fb},\;\;\;\;\sigma^{(2\gamma)}_R(e^+e^-\to\eta^\prime\gamma)
\approx 1.2\cdot 10^{-3}~\mathrm{fb}.
\label{eq4-10}\end{equation}
For $e^+e^-\to\eta\gamma$ the real part of the amplitude still dominates 
although not as drastically as for the constant $F_{\eta\gamma^*\gamma^*}$ 
form factor. For $e^+e^-\to\eta^\prime\gamma$ the real part is even somewhat 
smaller than the imaginary part. 

\section{Conclusions}
In this work we used a phenomenological, VMD inspired $F_{\eta\gamma^*
\gamma^*}$ form factor to estimate the expected contributions into the
$e^+e^-\to\eta\gamma$ and $e^+e^-\to\eta^\prime\gamma$ amplitudes from the
two-photon exchange diagrams of Fig.\ref{fig1-1}. The results are given in 
(\ref{eq3-35}) and (\ref{eq4-10}). The numbers obtained are too small 
to be of any importance for the recent BABAR analysis \cite{1-1}, but 
indicate that interference effects of the order of several percents are 
expected in future high precision studies of corresponding transition form 
factors.

Our results might be of some relevance for accurate evaluation of radiative 
corrections to  Dalitz decay $P\to e^+e^-\gamma$ \cite{3-15}, especially in 
light of some discrepancy with earlier studies \cite{1-3}, but we do not 
pursue this line of thought further in this article.  

\section*{Acknowledgments}
This investigation was initiated by V.~Druzhinin. The author thanks him for
valuable discussions. J.~D.~Jackson is acknowledged for sending his
unpublished report. The work is supported in part by grants 
Sci.School-905.2006.2 and RFBR 06-02-16192-a.

\end{document}